\documentclass[prb,endfloats,eqsecnum]{revtex4}
\usepackage{graphicx}
\newcommand{\e}{{\rm e}}
\newcommand{\be}{\begin{equation}}
\newcommand{\ee}{\end{equation}}
\newcommand{\ba}{\begin{eqnarray}}
\newcommand{\ea}{\end{eqnarray}}
\begin{document}
\title{Dynamical Multiple-Timestepping Methods for \\Overcoming
the Half-Period Time Step Barrier}

\author{Siu A. Chin}

\affiliation{Department of Physics, Texas A\&M University, 
College Station, TX 77843, USA}

\begin{abstract}
Current molecular dynamic simulations of biomolecules using multiple
time steps to update the slowingly changing force are hampered 
by an instability occuring at time step equal to half the period of the 
fastest vibrating mode. This has became a critical barrier preventing 
the long time simulation of biomolecular dynamics. Attemps to tame
this instability by altering the slowly changing force and efforts 
to damp out this instability by Langevin dynamics do not address 
the fundamental cause of this instability. In this work,
we trace the instability to the non-analytic character of the underlying 
spectrum and show that a correct splitting of the Hamiltonian, which
render the spectrum analytic, restores stability. The resulting Hamiltonian
dictates that in additional to updating the momentum due to the slowly 
changing force, one must also update the position with a modified mass.
Thus multiple-timestepping must be done dynamically. 
     
\end{abstract}
\maketitle

\section {Introduction}
The evolution of any dynamical variable $W(q_i,p_i)$ is given by the
Poisson bracket,
\begin{equation}
{{d }\over{dt}}W(q_i,p_i)=\{W,H\}\equiv
          \sum_i\Bigl(
		          {{\partial W}\over{\partial q_i}}
                  {{\partial H}\over{\partial p_i}}
				 -{{\partial W}\over{\partial p_i}}
                  {{\partial H}\over{\partial q_i}}
				                    \Bigr).
\label{peq}
\end{equation}
For the standard Hamiltonian, 
\begin{equation}
H(p,q)=\sum_i {p_i^2\over{2m_i}}+v(q_i),
\label{ham}
\end{equation}
the Poisson evolution equation (\ref{peq}) can be written as an operator
equation
\begin{equation}
{{dW }\over{dt}}=\sum_i\Bigl(
   {{p_i}\over{m_i}}{{\partial }\over{\partial q_i}}
              +F_i{{\partial }\over{\partial p_i}}
                  \Bigr)W,
\label{peqop}
\end{equation}
with formal solution
\begin{equation}
W(t)={\rm e}^{t(T+V)}W(0)=\Bigl[{\rm e}^{\epsilon (T+V)}\Bigr]^n W(0),
\label{peqform}
\end{equation}
where $T$ and $V$ are first order differential operators defined by
\begin{equation}
T\equiv\sum_i
   {{p_i}\over{m_i}}{{\partial }\over{\partial q_i}},\qquad
V\equiv\sum_i F_i{{\partial }\over{\partial p_i}}.
\label{tandv}
\end{equation}
Their exponentiations, ${\rm e}^{\epsilon T}$ and ${\rm e}^{\epsilon V}$, are then
{\it displacement} operators which displace $q_i$ and $p_i$ forward in time via  
\begin{equation}
q_i\rightarrow q_i+\epsilon{{p_i}\over{m_i}}\qquad{\rm and}\qquad 
p_i\rightarrow p_i+\epsilon F_i.
\label{peqv}
\end{equation}
Each factorization of ${\rm e}^{\epsilon (T+V)}$ into products  
of ${\rm e}^{\epsilon T}$, ${\rm e}^{\epsilon V}$ (and exponentials
of commutators of $T$ and $V$) give rises to a {\it symplectic} algorithm 
for evolving the system forward in time. This is the fundamental
Lie-Poisson theory of symplectic integrators, which has been studied 
extensively in the literature\cite{yos93,cha96,mcl02}. First and second order
factorizationa of the form
\ba
{\rm e}^{\epsilon (T+V)}
&&\approx {\rm e}^{\epsilon T}\e^{\epsilon V}\\
&&\approx {\rm e}^{{1\over 2}\epsilon V}{\rm e}^{\epsilon T}
{\rm e}^{{1\over 2}\epsilon V}
\label{symta}
\ea
give rises to the well-known symplectic Euler and the velocity-Verlet algorithm. 
Numerous higher order symplectic algorithms\cite{mcl02,for90,mcl91,mcl95,kos96,ome02}
are also known, but only a special class of fourth order algorithms can have 
strictly positive time steps as these lower order algorithms\cite{chin97,chin03}. 

In many cases, the Hamiltonian of interest is of the form,
\be
H(p,q)={p^2\over{2m}}+v_1(q)+v_2(q),
\label{twof} 
\ee
where there is a ``fast" force component $F_1=-\partial_q v_1$ and a
``slow'' force component $F_2=-\partial_q v_2$. For example, in biomolecular
dynamics, $F_1$ can be the rapidly vibrating force of H-O bonds and 
$F_2$, the sum of non-bond forces. Since it seems reasonable to sample the
slowly changing force $F_2$ less frequently, one can factorize this Hamiltonian
to first or second order,
\ba
\e^{\Delta t (T+V_1+V_2)}&&=
\e^{\Delta t (T+V_1)}
\e^{\Delta t V_2},\\
&&=\e^{{1\over 2}\Delta t V_2}
\e^{\Delta t (T+V_1)}
\e^{{1\over 2}\Delta tV_2},
\label{mts}
\ea
and solve for the fast force accurately using a smaller time step 
$\Delta\tau=\Delta t/k$,
\ba
\e^{\Delta t (T+V_1)}&&=
\Bigl[\e^{\Delta\tau T}
\e^{\Delta\tau V_1}\Bigr]^k,\\
&&=\Bigr[\e^{{1\over 2}{\Delta\tau}V_1}
\e^{\Delta\tau (T+V_1)}
\e^{{1\over 2}\Delta\tau V_1}\Bigr]^k.
\label{mtsp}
\ea
Thus the slow force is sampled at a multiple time steps of the fast force, 
$\Delta t=k \Delta\tau$. In the context of biomolecular simulation, this 
form of the multiple-time step (MTS) symplectic algorithm was 
introduced by Grubm\"uller {\it et al.}\,\cite{gru91},
and independently by Tuckerman {\it et al.}\,\cite{tuc92}. If a large time step $\Delta t$ 
can be used in MTS algorithms, one can hope to simulate the motion of marcomolecules 
through some biologically significant time intervals.

In the subsequent work of Zhou and Berne\cite{zho95} and
Watanabe and Karplus\cite{wat95}, this hope was dashed by the discovery of an
intransigent instability. No matter how accurately one has solved the fast force,
the MTS algorithm is unstable at $\Delta t=\pi/\omega_1$, where $\omega_1$ is 
the fast force's vibrational angular frequence. This has been described
as a ``resonance" instability\cite{bie93,man95,sch98}. However, the later numerical
work of Barth and Schlick\cite{bar98} clearly demonstrates that this instability 
exists at every mid-period as well, {\it i.e.}, at $\Delta t=n \pi/\omega_1=(n/2)T_1$, 
where $T_1$ is the period of the fast force, at $n=1,2,3$..., and not just 
at $n=2,4,6$,... Thus the notion of resonance is not a complete nor accurate
description of this instability. In this work, we 
will show that this instability is fundamentally 
related to the non-analytic character of the harmonic spectrum and cannot
be tamed by just multiple-timestepping the slow force. Stability can only be 
restored by a different splitting of the Hamiltonian requiring the slow 
force to be updated dynamically with a modified mass. 

In the next section, we analyze Barth and Schlick's model of MTS
instability\cite{bar98} and show that {\it static} multiple-timestepping
of the slow force destablizes the marginally-stable points of the fast force. 
In Section III, we show that an alternative splitting of 
the Hamiltonian, that of {\it dynamic} multiple-timestepping of the
slow force, restores stability. In Section IV, we explain why the particular
splitting worked in terms of the analytic character of the resulting spectrum.
Section V generalizes MTS to the case of multiple forces.
Section VI summarizes our findings and suggestions for large scale 
biomolecular simulations. 
          
\section {Stability analysis of MTS algorithms}

Barth and Schlick\cite{bar98} have proposed the simplest and clearest model 
for understanding the MTS instability. This is a harmonic oscillator with 
two spring constants,
\be
v_1(q)={1\over 2}\lambda_1 q^2, \quad v_2(q)={1\over 2}\lambda_2 q^2.
\nonumber
\ee 
Their numerical work unambiguously demonstrated the existence of MTS 
instability, but they did not carry their analysis far enough to pinpoint 
its origin. We will first complete their analysis of the symplectic Euler 
MTS algorithm.

Each operator $\e^{\Delta\tau T}$, $\e^{\Delta\tau V_1}$, when acting on
the canonical doublet $(p,q)$, produces a symplectic transformation, or map,
\begin{equation}
\left(\begin{array}{c}
       p^{n+1}\\
       q^{n+1}
      \end{array}\right)
= \e^{\Delta\tau V_1}
\left(\begin{array}{c}
       p^n\\
       q^n
       \end{array}\right)
= {\bf V}(\lambda_1,\Delta\tau)
\left(\begin{array}{c}
       p^n\\
       q^n
      \end{array}\right),
\label{veq}
\end{equation}
\begin{equation}
\left(\begin{array}{c}
       p^{n+1}\\
       q^{n+1}
      \end{array}\right)
= \e^{\Delta\tau T}
\left(\begin{array}{c}
       p^n\\
       q^n
      \end{array}\right)
= 
{\bf T}(m,\Delta\tau)
\left(\begin{array}{c}
       p^n\\
       q^n
      \end{array}\right),
\label{teq}
\end{equation}
where {\bf T} and {\bf V} are matrices given by
\ba
{\bf T}(m, \Delta \tau)&=&	
\left(\begin{array}{cc}
                 1 & 0 \\
                \Delta\tau/m & 1
      \end{array}\right),\label{tm}\\
{\bf V}(\lambda,\Delta \tau)&=& 
\left(\begin{array}{cc}
                 1 & -\Delta\tau \lambda \\
                 0 & 1
      \end{array}\right).
\label{vm}
\ea
The Jacobian of the transformation defined by
\be 
M={{\partial(p^{n+1},q^{n+1})}\over{\partial(p^n,q^n)} }
\nonumber
\ee
satisfies the defining symplectic condition
\be
M^T J M=J, \quad{\rm with}\quad 
  J=\left(\begin{array}{cc}
                 0 & -1 \\
                 1 & 0
      \end{array}\right), 
\ee
ensuring that det$M^T$ det$M$=1. For a sequence of symplectic
maps, by the chain-rule, the Jacobian multiplies
\be																
{{\partial(p_n,q_n)}\over{\partial(p_0,q_0)} }
={{\partial(p_n,q_n)}\over{\partial(p_{n-1},q_{n-1})} }...
{{\partial(p_2,q_2)}\over{\partial(p_1,q_1)} }
{{\partial(p_1,q_1)}\over{\partial(p_0,q_0)} }.
\ee
Regarding (\ref{veq},\ref{teq}) as numerical algorithms, the Jacobian 
matrix is just the error amplification matrix. However, only in the
present case of linear maps (\ref{veq},\ref{teq}) is the Jacobian
the same as the transformation matrix itself. 

The error amplification matrix 
corresponding to the symplectic Euler MTS algorithm 
\be
\e^{\Delta t (T+V_1+V_2)}= 
\Bigl[\e^{\Delta\tau T}
\e^{\Delta\tau V_1}\Bigr]^k
\e^{\Delta t V_2} +O(\Delta t^2) \\
\ee
is therefore (corresponding to Barth and Schlick's ${\bf A_I}$), 
\be
{\bf e}_E=
\Bigl[
{\bf T}(m, {{\Delta t}\over k})
{\bf V}(\lambda_1, {{\Delta t}\over k})
\Bigr]^k
{\bf V}(\lambda_2, \Delta t).
\label{error}
\ee
The symplectic matrices {\bf T} and {\bf V} as defined by 
(\ref{tm}) and (\ref{vm}), can also be expressed as exponentials 
of traceless matrices:
\ba
{\bf T}(m, \Delta \tau)&=&	
\exp\left[ \Delta \tau 
\left(\begin{array}{cc}
                 0 & 0 \\
                1/m & 0
      \end{array}\right)
\right],\\
{\bf V}(\lambda,\Delta \tau)&=&
\exp\left[ \Delta \tau 
\left(\begin{array}{cc}
                 0 & -\lambda \\
                 0 & 0
      \end{array}\right)
\right].
\label{expm}
\ea
For large multiple $k$, the fast force term in (\ref{error}) can be evaluated
analytically. Using the exponential forms for {\bf T} and {\bf V},
and invoking Trotter's theorem,
\ba
\lim_{k\rightarrow\infty}& 
\left(
\exp\left[ {{\Delta t}\over k} 
\left(\begin{array}{cc}
                 0 & 0 \\
                1/m & 0
      \end{array}\right)
\right]
\exp\left[{{\Delta t}\over k} 
\left(\begin{array}{cc}
                 0 & -\lambda_1 \\
                 0 & 0
      \end{array}\right)
\right]
\right)^k	   
=
\exp\left[\Delta t 
\left(\begin{array}{cc}
                 0 & -\lambda_1 \\
                 1/m & 0
      \end{array}\right),
\right]\nonumber\\
&=
\left(\begin{array}{cc}
                 \cos(\omega_1\Delta t) & -m\omega_1\sin(\omega_1\Delta t) \\
   (m\omega_1)^{-1}\sin(\omega_1\Delta t) & \cos(\omega_1\Delta t) 
      \end{array}\right)\equiv {\bf H}(m,\omega_1,\Delta t),  
\label{efinal}
\ea
where we have defined the fast force angular frequence $\omega_1=\sqrt{\lambda_1/m}$. 
Note that one starts with $\lambda_1$ and $m$, but the dynamics of the
system is governed by the square root $\omega_1$. Since $\omega_1$ is a non-analytic
function of $\lambda_1$ and $m$, it
can only be extracted in the limit of $k\rightarrow\infty$.  

The eigenvalues of the fast force 
error matrix (\ref{efinal}) is given by
\be
e_{1,2}=C\pm\sqrt{C^2-1}
\label{eigen}
\ee
with $C=\cos(x)$ and $x=\omega_1\Delta t$. 
The algorithm is marginally stable at all time step $\Delta t$ with 
$\vert e_{1,2}\vert=1$, but closest to being unstable at $x=n \pi$, 
where the two eigenvalues are degenerate, purely real, 
and equal to $\pm 1$.

The error matrix corresponding to Euler MTS algorithm (\ref{error}) 
is therefore 	
\be
{\bf e}_E= 
{\bf H}(m,\omega_1,\Delta t){\bf V}(\lambda_2,\Delta t).
\ee
The eigenvalues are still 
given by (\ref{eigen}), but now with $C$ altered to
\ba
C&&=\cos(x)-{1\over 2}\alpha x 
\sin(x),\label{cai}
\label{caih}\\
&&=A(x)\cos(x+\delta(x)),
\label{caii}
\ea
with $\alpha=\lambda_2/\lambda_1$, amplitude
\be
A(x)=\sqrt{1+(\alpha x/2)^2},
\label{ampe}
\ee 
and phase shift 
$\delta(x)=\tan^{-1}(\alpha x/2)$. The two C-functions, together
with the amplitude functions $\pm A(x)$, are plotted in Fig.1.
The Euler MTS algorithm is unstable whenever $|C(x)|>1$. 
As shown in Fig.1, the effect of 
$\lambda_2$, no matter how small, is to destablize marginally 
stable points of the fast force $\lambda_1$ at $x=n\pi$ into a finite band. The first band 
at $x=\pi$, is the half period barrier. The bands
are very narrow if $\lambda_2<<\lambda_1$. Within these instability 
bands, the extremes of the eigenvalues at $x+\delta(x)=n\pi$,
when $C=\pm A(x)$, are given by
(\ref{eigen}),
\be
e(x)=\pm\left(\alpha x/2+\sqrt{1+(\alpha x/2)^2}\right).
\label{mtse}
\ee
This is the linearly growing envelope of eigenvalues observed numerically
by Barth and Schlick\cite{bar98}. Since the eigenvalue departs from unity 
linearly as a function of $x$, we can characterize this
instability as first order in $x$. This is the most important characterization 
of MTS algorithms and is plotted in Fig. 2. As one can see, as long as $\alpha$ is 
not zero, the departure from unity will be significant at $x=\pi$, which 
explains the persistence of the half period barrier. We emphasize that
$e(x)$ only gives the correct eigenvalues at $x+\delta(x)=n\pi$,
when $C=\pm A(x)$. For $\alpha<<1$, this means that $e(x)$ is only
correct at $x\approx n\pi$. For other values of $x$, $e(x)$ is not 
the correct eigenvalue and the algorithm is actually stable.

The error matrix for the second order Verlet-like MTS algorithm,
\be
{\bf e}_V= {\bf V}(\lambda_2,{1\over 2}\Delta t)\,
      {\bf H}(m,\omega_1,\Delta t) 	
	  {\bf V}(\lambda_2,{1\over 2}\Delta t)
\ee
has the same C-function (\ref{cai}) and therefore the identical 
first order instability problem. This is a surprise. As we will see later in Section IV, 
increasing the order of {\it static} MTS algorithms does little to increase 
its stability.

\section {Restoring Stability via Dynamical MTS}
The MTS algorithm in the last section splits the 
Hamiltonian as 
\be
H(p,q)=\left({p^2\over{2m}}+{1\over 2}\lambda_1 q^2\right)
+{1\over 2}\lambda_2 q^2,
\label{twosp} 
\ee												 
where the parenthesis describes the full dynamics of spring $\lambda_1$.
This leaves $\lambda_2$ as only a static force with 
no direct role in changing the particle's position. We shall refer to this
as {\it static} multiple-timestepping. This is not an equitable 
splitting, nor the only one possible. The Hamiltonian can alternatively
be splitted as	 
\be
H(p,q)=\left({p^2\over{2m_1}}+{1\over 2}\lambda_1 q^2\right)
+\left({p^2\over{2m_2}}+{1\over 2}\lambda_2 q^2\right),
\label{twospp} 
\ee
with the constraint
\be
{1\over m_1}+{1\over m_2}={1\over m}.
\label{mass}
\ee
Now both springs are fully dynamical and we can use the freedom in the choice
of $m_1$ and $m_2$ to maximize stability. We shall refer to this
as {\it dynamic} multiple-timestepping. The Euler splitting algorithm of 
(\ref{twospp}) in operator form is
\be
\e^{\Delta t (T_1+V_1+T_2+V_2)}\approx
\e^{\Delta t (T_1+V_1)}
\e^{\Delta t (T_2+V_2)}.
\label{dmts}
\ee
Consider first when both are evaluated exactly as in (\ref{efinal}), 
then the error matrix is
\be
{\bf e}_{DE}=
      {\bf H}(m_1,\Omega_1,\Delta t) 	
      {\bf H}(m_2,\Omega_2,\Delta t),
\label{dmtse}	   		 
\ee
with 
\be
\Omega_1=\sqrt{\lambda_1\over m_1}\quad{\rm and}\quad
\Omega_2=\sqrt{\lambda_2\over m_2}.
\label{bigw}
\ee\
The corresponding C-function is
\be
C=\cos((\Omega_1+\Omega_2)\Delta t)
-{{(m_1\Omega_1-m_2\Omega_2)^2}\over{2 m_1\Omega_1 m_2\Omega_2}}
\sin(\Omega_1\Delta t)
\sin(\Omega_2\Delta t).
\label{dmtsc}
\ee
The destablizing sine function term can be eliminated by
choosing
\be
m_1\Omega_1=m_2\Omega_2\quad\rightarrow \quad m_1\lambda_1=m_2\lambda_2.
\label{meq}
\ee
Thus stability can be fully restored in this splitting with
the choice of
\be
{1\over m_1}={\lambda_1\over{\lambda_1+\lambda_2}}{1\over m}
\quad{\rm and}\quad 
{1\over m_2}={\lambda_2\over{\lambda_1+\lambda_2}}{1\over m}.
\label{mval}
\ee
For this choice of $m_1$ and $m_2$, we observe that
\be
\Omega_1={\lambda_1\over{\lambda_1+\lambda_2}}\Omega 
\quad{\rm and}\quad
\Omega_2={\lambda_2\over{\lambda_1+\lambda_2}}\Omega 
\ee
where
\be
\Omega=\sqrt{ {\lambda_1+\lambda_2}\over{m}}
\ee
is the exact angular frequence of the system. This means, however
that
\be
\Omega=\Omega_1+\Omega_2,
\ee
{\it i.e.}, the choice of $m_1$ and $m_2$ which restores stability 
also linearizes the spectrum. To compare with the static case, we also
note that
\be
\Omega_1=\sqrt{{\lambda_1\over{\lambda_1+\lambda_2}}{\lambda_1\over m}}
={\omega_1\over\sqrt{1+\alpha}}
\ee
and
\be
\Omega_2=\alpha\,\Omega_1.
\ee

For MTS algorithms, we do not want to evaluate the second 
spring force exactly, but only sparingly. Thus we further
approximate (\ref{dmts}) by
\be
\e^{\Delta t (T_1+V_1+T_2+V_2)}\approx
\e^{\Delta t (T_1+V_1)}
\e^{\Delta t T_2}\e^{\Delta t V_2}.
\ee
This is the dynamical Euler MTS algorithm with error matrix
\be
{\bf e}_{DE}=
      {\bf H}(m_1,\Omega_1,\Delta t)
	  {\bf T}(m_2,\Delta t)
	  {\bf V}(\lambda_2,\Delta t).
\label{dmtsex}
\ee
The resulting C-function is
\be
C= \cos(x^\prime)(1-{1\over 2}(\alpha x^\prime)^2)-\alpha x^\prime \sin(x^\prime),
\label{cdmte}
\ee
where 
\be
x^\prime=\Omega_1\Delta t=x/\sqrt{1+\alpha}\quad{\rm and}\quad 
\alpha x^\prime=\Omega_2\Delta t.
\ee 
This C-function is 
$\cos(\Omega_1\Delta t+\Omega_2\Delta t)$
correct to second order in $\Omega_2\Delta t$. The corresponding
amplitude and eigenvalue functions are
\be
A(x^\prime)=\sqrt{1+(\alpha x^\prime)^4/4},
\label{ampp}
\ee
\be
e(x^\prime)=\pm\left[\,(\alpha x^\prime)^2/2
+\sqrt{1+(\alpha x^\prime)^4/4}\,\right].
\label{edmts}
\ee
Thus by allowing $\lambda_2$ to be dynamical, the same effort in force
evaluation improves the instability to second order. This is shown in
Fig.2. However, one can do even better. By (\ref{dmtse}), the algorithm's 
stability will continue to improve with improvements in solving $\lambda_2$'s 
dynamics. With still only one slow force evaluation, one can solve $\lambda_2$'s 
dynamic to second order with error matrix
\be
{\bf e}_{DE2}=
      {\bf H}(m_1,\Omega_1,\Delta t)
	  {\bf T}(m_2,{1\over 2}\Delta t)
	  {\bf V}(\lambda_2,\Delta t)
	  {\bf T}(m_2,{1\over 2}\Delta t),
\label{dmtsep}
\ee
C-function
\be
C= \cos(x^\prime)(1-{1\over 2}(\alpha x^\prime)^2)
-\alpha x^\prime(1-{1\over 8}(\alpha x^\prime)^2) \sin(x^\prime),
\label{cdmtep}
\ee
amplitude 
\be
A(x^\prime)=\sqrt{1+(\alpha x^\prime/2)^6},
\label{amppp}
\ee
eigenvalue
\be
e(x^\prime)=\pm\left[\,(\alpha x^\prime/2)^3
+\sqrt{1+(\alpha x^\prime/2)^6}\,\right],
\label{edmtsx}
\ee
and improve stability to third order! In sharp contrast to the static 
case, where the use of a second order algorithm for the slow force 
yielded no improvement in stablity, the improvement here is dramatic.
As shown in Fig.2, even for $\alpha$ as large as 1/20, this second order 
algorithm is basically stable at $x=\pi$.

If one is willing to evaluate the slow force more than
once, further systematic improvments are possible. The second spring's 
motion can be solve to fourth order 
using forward symplectic algorithm 4A \cite{chin97,chin03}: 
\be
\e^{\Delta t (T_2+V_2)}=
\e^{{1\over 2}\Delta t V_2}
\e^{{1\over 2}\Delta t T_2}
\e^{{2\over 3}\Delta t \widetilde V_2}
\e^{{1\over 2}\Delta t T_2}
\e^{{1\over 6}\Delta t V_2}+O(\Delta t)^5.
\ee
Here $\widetilde V_2=V_2+{1\over 48}\Delta t^2[ V_2,[T_2,V_2]]$.
The double commutator modifies the original spring constant
$\lambda_2$ to 
\be
\widetilde\lambda_2
=
\lambda_2(1-{1\over {24}}{\lambda_2\over m_2}\Delta t^2)
=\lambda_2(1-{1\over {24}}(\alpha x^\prime)^2).
\label{springt}
\ee
The resulting error matrix is
\be
{\bf e}_{4A}= {\bf H}(m_1,\Omega_1,\Delta t)
              {\bf V}(\lambda_2,{1\over 6}\Delta t)
              {\bf T}(m_2,{1\over 2}\Delta t) 	
	          {\bf V}(\widetilde\lambda_2, {2\over 3}\Delta t)
              {\bf T}(m_2,{1\over 2}\Delta t) 	
              {\bf V}(\lambda_2,{1\over 6}\Delta t),
\label{efoura}
\ee
with C-function
\ba
C&=&\cos(x^\prime)\left[1-{1\over 2}(\alpha x^\prime)^2
+{1\over {24}}(\alpha x^\prime)^4
-{1\over {864}}(\alpha x^\prime)^6\right]\nonumber\\
&&-\left[
\alpha x^\prime -{1\over 6}(\alpha x^\prime)^3+{7\over {864}}(\alpha x^\prime)^5
-{1\over {10368}}(\alpha x^\prime)^7\right] \sin(x^\prime),
\label{cdmteppp}
\ea
amplitude 
\be
A(x^\prime)=\sqrt{
1
+{1\over {3^6}}(\alpha x^\prime/2)^{10}
-{2\over {3^7}}(\alpha x^\prime/2)^{12} 
+{1\over {3^8}}(\alpha x^\prime/2)^{14} },
\label{ampppp}
\ee
and eigenvalue function,
\be
e(x^\prime)= \pm\left[ \sqrt{A^2(x^\prime)-1}+A(x^\prime)
\right].
\label{epppp}
\ee
The instability is now pushed back to fifth order in $x$. Fig. 2 shows
that even for $\alpha$ as large as 1/20, this algorithm is now basically stable 
out to $x\approx 6\pi$. For $\alpha=1/400$, as considered by Barth and Schlick, 
this algorithm has $e\leq 1.00001$ at $x\approx 50 \pi$. There is no doubt that 
one has overcame the half-period barrier at $x=\pi$.
 
\section{Stability Explained}

      The poor stability of static multiple-timestepping can be traced
to the non-analytic character the spectrum. The system's exact angular frequence is
\be
\Omega=\sqrt{ {{\lambda_1+\lambda_2}\over{m}} }
=\sqrt{\omega_1^2+\omega_2^2}
=\omega_1\sqrt{1+\alpha},
\label{wst}
\ee
with exact C-function
\be
C=\cos(\Omega\Delta t).
\label{exactc}
\ee
In terms of $x=\omega_1\Delta t=\sqrt{\lambda_1/m}\Delta t$ and 
$\alpha=\lambda_2/\lambda_1$, this function is non-analytic in
$\alpha$,
\be
C=\cos(x\sqrt{1+\alpha}).
\label{cst}
\ee
When expanded in terms of $\alpha$, it has the form
\be
C=\cos(x)-{1\over 2}x \sin(x) \alpha +
\left[ {1\over 8} x \sin(x)-{1\over 8}x^2 \cos(x)\right]\alpha^2+\cdots.
\label{expand}
\ee
The first order term is precisely the first order result (\ref{caih}). If one
were able to reproduce this expansion, one could in principle systematically
restore stability. Unfortunately one cannot; when regarding $\lambda_2$ as 
static, one must expand in powers of 
${\bf V}(\lambda_2,\Delta t)\propto \lambda_2\Delta t\propto\alpha x$, and can
never reproduce the term $\propto x\alpha^2$ in (\ref{expand}) in any finite order. 
Worse, second and fourth order algorithms do not even reproduce the $(x\alpha)^2$ 
term with the correct coefficient. 

By contrast, in dynamical multiple-timestepping, one has,
\be
\Omega=\Omega_1+\Omega_2,
\label{wdyn}
\ee
and the spectrum is linear in $\Omega_2$. The corresponding C-function
\be
C=\cos(\Omega_1\Delta t+\Omega_2\Delta t)=\cos(x^\prime+\alpha x^\prime),
\label{cdyn}
\ee
as shown in the last section, can be systematically reproduced order 
by order in $(\alpha x^\prime)$. Thus dynamical multiple-timestepping 
linearizes the spectrum and can overcome the half period barrier by
going to higher order.

\section{Generalization to Many Forces}

For more than two forces, the generalization is easy. Again, using
the harmonic oscillator as an illustration, the ``$N$-forces" case of
\be
H(p,q)={p^2\over{2m}}+{1\over 2}\sum_{i=1}^N\lambda_i q^2,
\label{threef} 
\ee
can be dynamically splitted as
\be
H(p,q)=\sum_{i=1}^N \left({p^2\over{2m_i}}+{1\over 2}\lambda_i q^2\right),
\label{threeh} 
\ee
with the primary constraint
\be
\sum_{i=1}^N{1\over m_i}={1\over m}.
\label{nmass}
\ee
and the pair-wise stability conditions, $i\neq j$,
\be
m_i\lambda_i=m_j\lambda_j.
\label{nstable}
\ee
Both can be easily satisfied by the following generalization of 
(\ref{mval}),
\be
{1\over m_i}={\lambda_i\over{\sum_{j=1}^N\lambda_j}}{1\over m}
=\left({\omega_i\over{\Omega}}\right)^2{1\over m}.
\ee
Thus the inverse of the dynamical mass should be chosen in proportional
to the strength of the force, or the square of its angular frequence.

\section{Conclusions}

In this work, we have given a detailed analysis of Barth and Schlick's 
model of MTS instability\cite{bar98}. We show that the instability of static
MTS algorithms can ultimately be traced to the non-analytic character of the 
underlying spectrum. Static MTS algorithms are simply very
poor starting points for solving such a spectrum, even if one were to modify
or average over the slow force\cite{lza99}. By contrast, dynamic
MTS algorithms linearize the spectrum, render it analytic, and can improve
stability systematically order by order. The use of Langevin 
dynamics to damp out the instability\cite{bar982} simply masks the true 
dynamics of the system without fundamentally solving the instability
problem.
 
Realistic biomolecular simulations are too complicated for a detailed stability
analysis as in the harmonic oscillator case. Nevertheless, the harmonic oscillator 
captures the essence of its fast vibrating modes. Thus the insight of
dynamic multiple-timestepping can be applied easily. The key idea is to
decompose 
\be
{1\over m}={1\over m_1}+{1\over m_2}
\label{last}
\ee
and update particles affected by the slow force dynamically with mass $m_2$. 
In the harmonic oscillator case, $1/m_1$ and $1/m_2$ are to be determined 
in proportional to the strength, or the square of the frequence, of the force. 
For realistic simulations, one can simply determine the optimal $m_2$ by 
trial-and-error subject to the constraint (\ref{last}).

\begin{acknowledgments}
This work was supported, in part, by the National Science Foundation
grants No. PHY-0100839 and DMS-0310580. 

\end{acknowledgments}

\begin{figure}
	\vspace{0.5truein}
	\centerline{\includegraphics[width=0.8\linewidth]{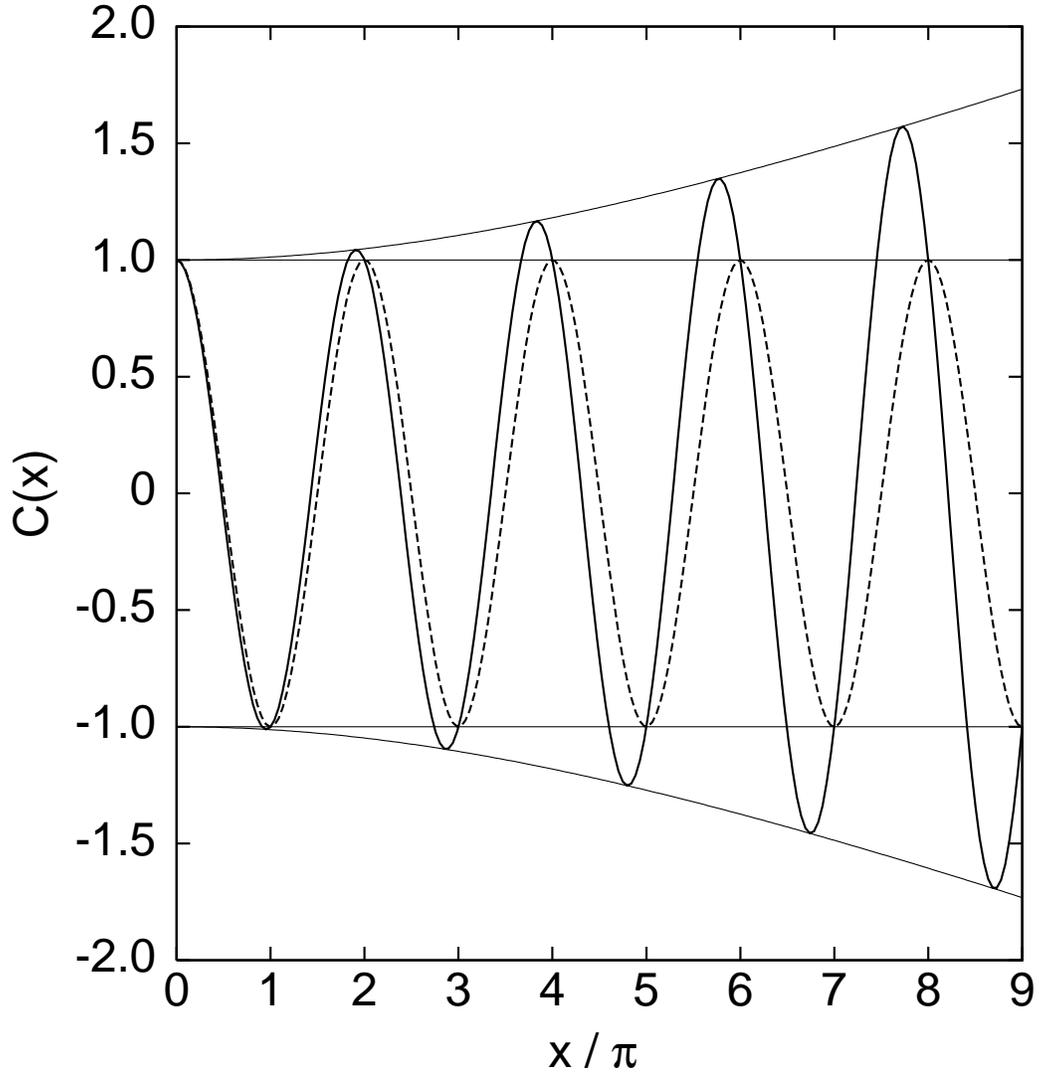}}
	\vspace{0.5truein}
\caption{ The fundamental C-function for determing the stability
of Multiple-Timestepping (MTS) algorithms.  The dashed line gives the 
stable C-function for the fast force alone, $C(x)=\cos(x)$ 
where $x=\omega_1\Delta t$, and $\omega_1$ is the vibrational
angular frequence of the fast force. The solid lines give the
C-function for the static Euler MTS algorithm, Eq.(\ref{cai}).
To make the unstable regions visible, a large value
of $\alpha=\lambda_2/\lambda_1=1/10$ is used, where $\lambda_1$
and $\lambda_2$ are the force constant of the fast and slow force
respectively. The algorithm is unstable whenever $|C(x)|>1$.
The most unstable point in each unstable band near $x\approx n\pi$ 
touches the amplitude envelope $\pm A(x)$, Eq.(\ref{ampe}). 
\label{fig1}}
\end{figure}
\begin{figure}
	\vspace{0.5truein}
	\centerline{\includegraphics[width=0.8\linewidth]{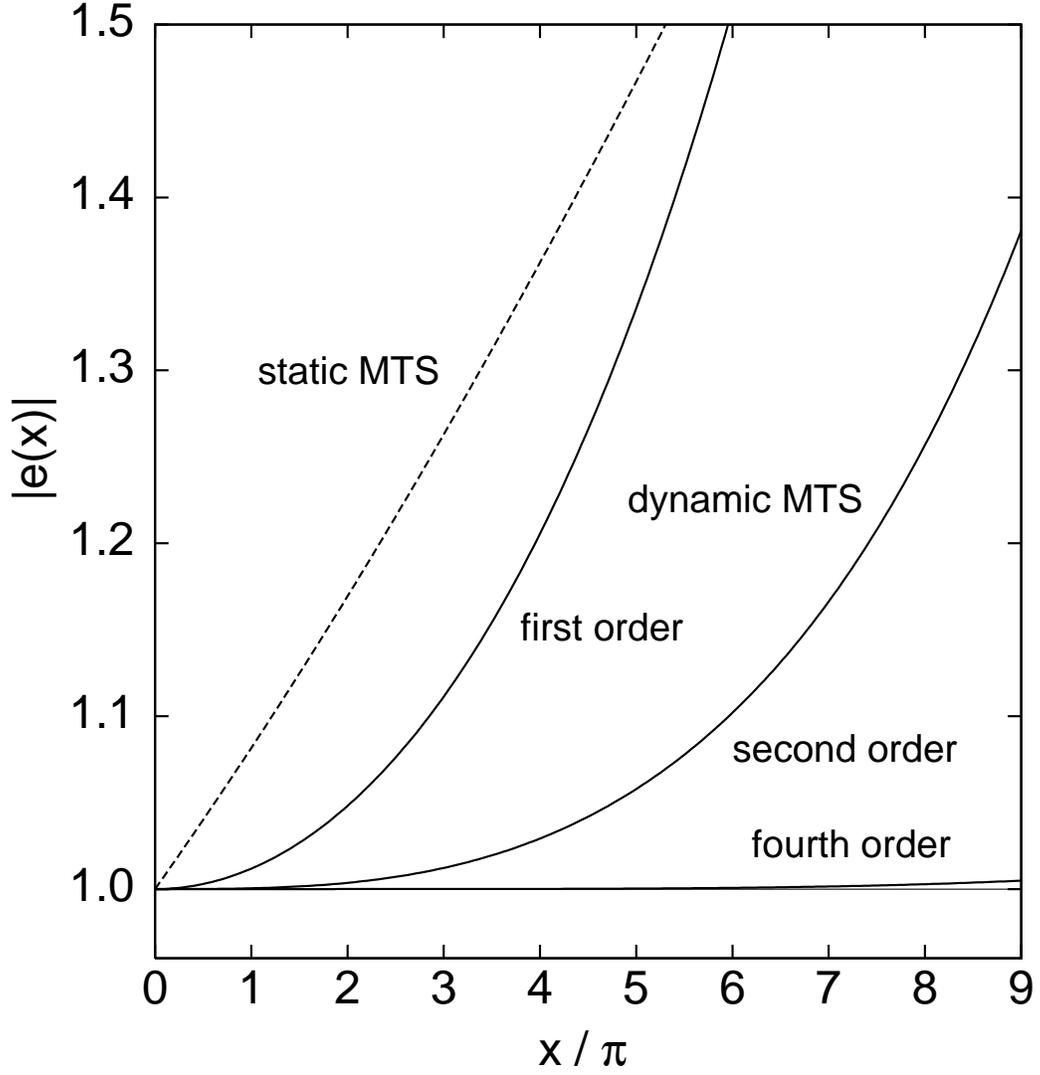}}
	\vspace{0.5truein}
\caption{The magnitude of the error matrix's eigenvalue
 for various MTS algorithms. The dashed line is the static Euler MTS
 algorithm. The three solids lines are the three dynamic MTS algorithms
 described in the text. The algorithm is unstable whenever $|e(x)|>1$,
 however, in this graph, only values at $x+\delta(x)= n\pi$ are
 true eigenvalues. See text for details. A large value of $\alpha=1/20$
 is used to make the fourth order dynamic MTS result visible.  
\label{fig2}}
\end{figure}

\end{document}